\documentclass[prl,showpacs,twocolumn,superscriptaddress]{revtex4}

\usepackage{amsfonts,amssymb,amsmath}
\usepackage{mathtools}
\usepackage[]{graphics,graphicx,epsfig}
\usepackage{amsthm,multirow}
\begin{document}

\title{Entropic tests of multipartite nonlocality and state-independent contextuality}

\author{Sadegh \surname{Raeisi}}
\email{sraeisi@uwaterloo.ca}
\affiliation{Institute for Quantum Computing, University of Waterloo, Ontario, Canada}
\affiliation{Centre for Quantum Technologies, National University of Singapore, 3 Science Drive 2, 117543 Singapore, Singapore}

\author{Pawe{\l} \surname{Kurzy\'nski}}
\email{cqtpkk@nus.edu.sg}
\affiliation{Centre for Quantum Technologies, National University of Singapore, 3 Science Drive 2, 117543 Singapore, Singapore}
\affiliation{Faculty of Physics, Adam Mickiewicz University, Umultowska 85, 61-614 Pozna\'{n}, Poland}

\author{Dagomir \surname{Kaszlikowski}}
\email{phykd@nus.edu.sg}
\affiliation{Centre for Quantum Technologies, National University of Singapore, 3 Science Drive 2, 117543 Singapore, Singapore}
\affiliation{Department of Physics, National University of Singapore, 2 Science Drive 3, 117542 Singapore, Singapore}

\date{\today}

%%%%%%%%%%%%%%%%%%%%%%%%%%%%%%%%%%%%%%%%%%%%%%%%%%

\begin{abstract}

We introduce a multipartite extension of an information-theoretic distance first introduced in [Nature 341, 119 (1989)]. We use this new distance to derive entropic tests of multipartite nonlocality for three and for an arbitrary even number of qubits as well as a test of state-independent contextuality. In addition, we re-derive the tripartite Mermin inequality and a state-independent non-contextuality inequality by Cabello [Phys. Rev. Lett. 101, 210401 (2008)]. This suggests that the information-theoretic distance approach to multipartite nonlocality and state-independent contextuality can provide a more general treatment of nonclassical correlations than the orthodox approach based on correlation functions.

\end{abstract}

\maketitle

%%%%%%%%%%%%%%%%%%%%%%%%%%%%%%%%%%%%%%%%%%%%%%%%%%

\begin{figure}[t]
    \begin{center}
    	\includegraphics[width=1.0\columnwidth,trim=4 4 4 4,clip]{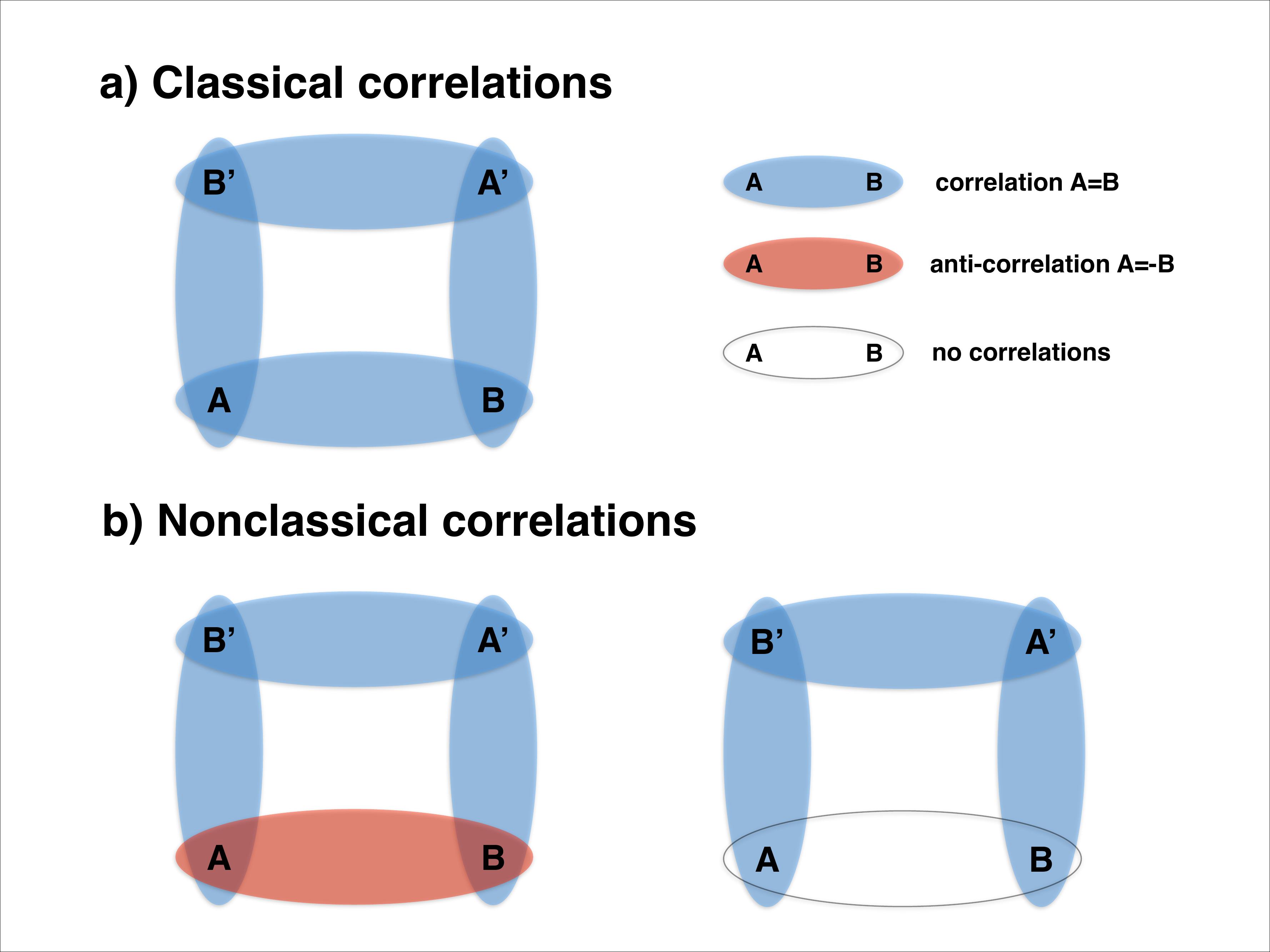}
        \caption{Bipartite correlations between two observers. Each local measurement is maximally random, i.e., $P(A=1)=P(A=-1)=1/2$, etc. Classical correlations (\emph{a}) are transitive, namely if $A=B'$, $B'=A'$ and $A'=B$, then $A=B$. However, nonclassical correlations can be non transitive giving $A=-B$ (\emph{b} left), or no correlations $H(AB)=H(A)+H(B)$ (\emph{b} right). Such extreme nonclassical bipartite correlations cannot be observed between quantum systems.}
        \label{fig1}
    \end{center}
\end{figure}

\emph{Introduction.} In classical information theory if some binary property $A$ is correlated with $B'$, $B'$ with $A'$ and $A'$ with $B$ then $A$ must be correlated with $B$. This is not necessarily true in non-classical information theories where correlations can be non-transitive. For instance, $A$ can be anti-correlated with $B$ \cite{PRbox} (see Fig. \ref{fig1}). If one looks only at the outcomes of random variables, the classical and nonclassical scenarios are dramatically different, however from the entropic point of view they do not differ at all \cite{Chaves}. More precisely, the Shannon entropies $H(A)$, $H(B)$ and $H(AB)$, where  $H(A)=-\sum_{a}P(A=a)\log_2 P(A=a)$, are the same regardless of whether the system is classical or not. 

In order to detect nonclassicality via entropic test one has to look for other types of nonclassical correlations (see Fig. \ref{fig1} b right). These can be found by either looking for a different set of measurements \cite{BC}, or by a post-processing of a measured data, e.g., mixing of nonclassical and classical distributions \cite{Chaves}. 

The entropic tests of the bipartite nonlocality were introduced in the late 80s \cite{BC}. Recently, these tests were extended to include the state-dependent contextuality \cite{entropic,CF1,CF2}, which is a more general concept than nonlocality since it studies nonclassical correlations in systems that are not necessarily spatially separated. However, the entropic approach to the multipartite nonlocality and to the state-independent contextuality has not been proposed before. 

We use the previously developed information-theoretic distance approach to nonclassical correlations \cite{Santos,distance1,distance2} and propose a new multipartite distance that can be applied to binary $\pm 1$ measurements. This distance quantifies multipartite correlations in terms of Shannon entropy and we apply it to derive an entropic tripartite inequality. The structure of this inequality resembles the tripartite Mermin inequality \cite{Mermin}. Next, we derive an entropic multipartite inequality to test the multipartite nonlocality of an arbitrary even number of qubits. Finally, we derive an entropic inequality to test the state-independent contextuality, which resembles the correlation-based inequality by Cabello \cite{Cabello}. All these inequalities are satisfied by correlations for which the information-theoretic distance can be properly defined. This is true for local realistic and non-contextual systems, but is violated by measurements on quantum systems.

%%%%%%%%%%%%%%%%%%%%%%%%%%%%%%%%%%%%%%%%%%%%%%%%%%

\emph{Multipartite distance.} The entropy-based information-theoretic distance was originally proposed by Zurek \cite{Zurek}. The essence of distance is to measure how far {\it two points} are. Points can be represented by coordinates in a Cartesian space or by a more abstract objects such as functions or operators. In the information-theoretic framework these objects are two random variables -- in our case, two jointly measurable observables. One of the obvious questions to ask is whether it is possible to extend the notion of distance to more than two points. Here, we show that this can be done if points are represented by binary, jointly measurable random variables, however, there are other ways to propose the multipartite information-theoretic distance (see for example \cite{Vitanyi}).

Let us first consider the following function defined for binary observables $A$ and $B$
\begin{equation}
d\left(A,B\right)=H\left(A \cdot B\right),\label{eq:Distance}
\end{equation}
The measurement of $A\cdot B$ is one where the outcomes are the product $ab$. 

The function in Eq.(\ref{eq:Distance}) satisfies all of the distance
properties. It is (i) \emph{non-negative}, $d\left(A,B\right)\geq0$, because $H\left(X\right)\geq 0$ and it equals to zero only if $A=B$, in which case the outcome of the measurement is always one (ii) \emph{symmetric} $d\left(A,B\right)=d\left(B,A\right)$ (iii) it obeys \emph{triangle inequality} $H\left(A\cdot B\right)\leq H\left(B \cdot C\right)+H\left(A \cdot C\right).$

The triangle inequality is satisfied because $H\left(A\cdot B|A\cdot C,B\cdot C\right) = 0$, i.e. if the outcomes of the two measurements $A\cdot C$ and $B\cdot C$ are known,
then the outcome of $A\cdot B$ is the product of the two outcomes and is therefore known. More precisely, $H\left(A\cdot B\right)\leq H\left(A\cdot B,B \cdot C,A\cdot C\right)=H\left(A\cdot B|A\cdot C,B\cdot C\right)+H\left(B\cdot C,A\cdot C\right)=H\left(B\cdot C,A\cdot C\right)\leq H\left(B\cdot C\right)+H\left(A\cdot C\right).$ We used $H(AB)=H(A|B) + H(B)$, $H(AB) \leq H(A) + H(B)$ and $H(A) \leq H(AB)$.

The distance in Eq. (\ref{eq:Distance}) can be extended to multipartite measurements. Note, that for
a set of binary $\pm 1$ variables $\left\{ A_{1},A_{2},\dots,A_{n}\right\} $ one can define 
\begin{equation}\label{d}
\delta \left(A_{1},A_{2},\dots,A_{n}\right)=H\left(A_{1}\cdot A_{2} \cdot \ldots \cdot A_{n}\right),
\end{equation}
which is the natural extension of the distance for two variables. 

The function $\delta$ is obviously non-negative and symmetric, but it also has a nice associative property
\begin{eqnarray}
& &\delta \left( A_{1},\ldots,A_{k},A_{k+1},\ldots,A_n \right)= \nonumber \\ & &\delta\left( \left( A_{1}\cdot\ldots\cdot A_{k}\right),\left(A_{k+1}\cdot\ldots\cdot A_{n}\right)\right). \label{eq:AssociativeProperty}
\end{eqnarray}
Note that using the symmetry property, any two $A_{i}$ could be
associated. Moreover, the associativity also implies that $\delta$ obeys the following version of the triangle inequality 
\begin{eqnarray}
& &\delta \left(A_{1},\ldots,A_{k},A_{k+1},\ldots,A_n \right) = \nonumber \\ 
& &\delta \left((A_{1}\cdot\ldots\cdot A_{k}),(A_{k+1}\cdot\ldots\cdot A_n) \right) \leq \nonumber \\
& & \delta \left((A_{1}\cdot\ldots\cdot A_{k})\cdot(B_{1}\cdot\ldots\cdot B_{m})\right) + \nonumber \\
& & \delta\left((B_{1}\cdot\ldots\cdot B_{m})\cdot(A_{k+1}\cdot\ldots\cdot A_n) \right) = \nonumber \\
& & \delta \left(A_{1},\ldots, A_{k},B_{1},\ldots,B_{m}\right) + \nonumber \\
& & \delta\left(B_{1},\ldots, B_{m},A_{k+1},\ldots, A_n \right). \label{triangle}
\end{eqnarray}

We would like to remark that multipartite information diastance has been considered before. In Ref. \cite{Vitanyi} Vitanyi considered the following quantity
\begin{equation}
E_{\max}(X) = \max_{x\in X} K(X|x),
\end{equation}
where $X$ is a set, $x$ are its elements and $K$ stands for Kolmogorov complexity. However, $E_{\max}(X)$ with $K$ replaced by Shannon entropy $H$ cannot be used to detect the difference between classical and nonclassical correlations. This motivated us to look for Eq. (\ref{d}).

%%%%%%%%%%%%%%%%%%%%%%%%%%%%%%%%%%%%%%%%%%%%%%%%%%

\emph{Tripartite information-theoretic Bell inequality.} Let us examine the properties of (\ref{d}) in the context of tripartite measurements. We derive the following inequality: 
\begin{eqnarray}
& &\delta \left(A_{1},B_{1},C_{1}\right)\leq d\left(A_{1},\left(B_{2}.C_{2}\right)\right)+d\left(\left(B_{2}.C_{2}\right),\left(B_{1}.C_{1}\right)\right) \nonumber \\
& &=  d\left(A_{1},B_{2}.C_{2}\right)+\delta \left(B_{2},C_{1},B_{1},C_{2}\right) \nonumber \\
& & \leq   \delta \left(A_{1},B_{2},C_{2}\right)+d\left(A_{2},B_{2}\cdot C_{1}\right)+d\left(A_{2},B_{1}\cdot C_{2}\right) \nonumber \\
& &=  \delta \left(A_{1},B_{2},C_{2}\right)+\delta \left(A_{2},B_{2},C_{1}\right)+\delta \left(A_{2},B_{1},C_{2}\right). \label{eq:EntropicMermin}
\end{eqnarray}
where $A_i$, $B_j$ and $C_{k}$ ($i,j,k=1,2$) are measurements of Alice, Bob and Charlie, respectively.

\begin{figure}[t]
    \begin{center}
    	\includegraphics[width=1.0\columnwidth,trim=4 4 4 4,clip]{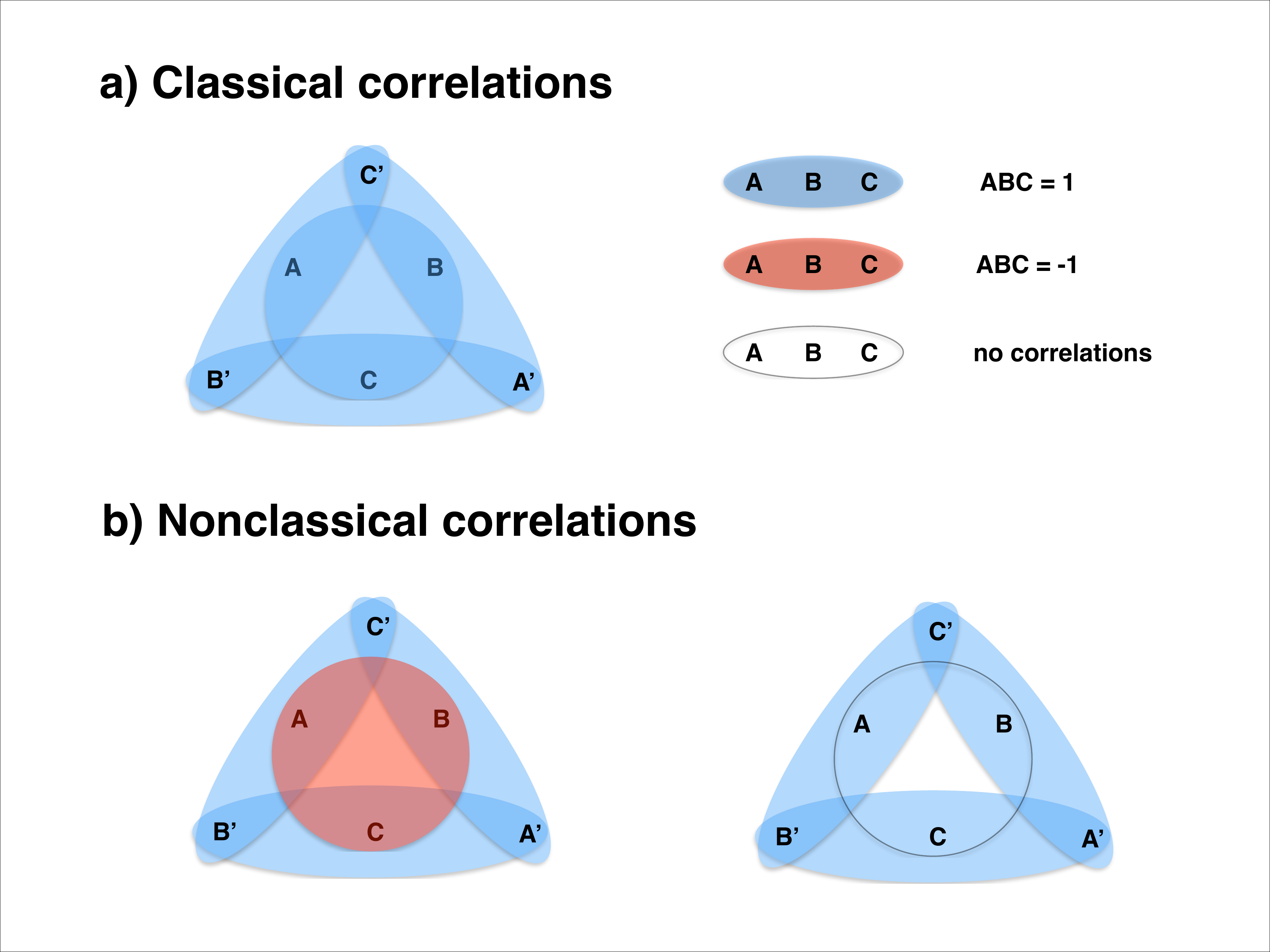}
        \caption{Tripartite correlations between three observers. Each local measurement is maximally random, i.e., $P(A=1)=P(A=-1)=1/2$, etc. Classical correlations (\emph{a}) imply that if $AB'C'=A'BC'=A'B'C=1$, then $ABC=1$. However, for nonclassical tripartite correlations (\emph{b} left) one can observe $ABC=-1$ (GHZ paradox). The above nonclassical correlations are entropically indistinguishable from the classical ones, however one can also observe no tripartite correlations between $A$, $B$ and $C$ (\emph{b} right). These can be detected via entropic inequalities, since the product of the three observables has maximal entropy $H(A\cdot B\cdot C)=1$. In contrast to bipartite scenario, such extreme nonclassical tripartite correlations can be observed between quantum systems.}
        \label{fig2}
    \end{center}
\end{figure}

The inequality (\ref{eq:EntropicMermin}) was derived using the classical properties of Shannon entropy, therefore it must hold in any theory that obeys them. In particular, in local realistic theories there exists a joint probability distribution for all observables $A_1,\dots,C_2$ \cite{Fine} and as a consequence there exists a joint entropy $H(A_1\cdot \ldots \cdot C_2)$ which implies the validity of (\ref{eq:EntropicMermin}). However, one may expect the violation of this inequality in a nonclassical theory (see Fig. \ref{fig2}).

Let us consider a three-qubit system in the Greenberger-Horne-Zeilinger (GHZ) state $|GHZ\rangle=\frac{1}{\sqrt{2}}(|000\rangle + |111\rangle)$ shared between Alice, Bob and Charlie. Each of them performs one of the two possible local $\pm1$ measurements on their subsystem: $A_1,A_2,B_1,\dots$ As previously discussed, we can choose $\delta(A_i,B_j,C_k)=H(A_i \otimes B_j \otimes C_k)$ (for $i,j,k=1,2$) and plug these measurements to the inequality (\ref{eq:EntropicMermin}) to obtain 
\begin{eqnarray}
& & H\left(A_{1}\otimes B_{1} \otimes C_{1}\right) \leq H\left(A_{1}\otimes B_{2} \otimes C_{2}\right) \nonumber \\&+& H\left(A_{2}\otimes B_{1} \otimes C_{2}\right) + H\left(A_{2}\otimes B_{2} \otimes C_{1}\right). \label{quantum}
\end{eqnarray}

Quantum theory violates the inequality (\ref{quantum}) if Alice, Bob and Charlie chose
\begin{eqnarray}
A_{1}=B_{1}=C_{1} & = & \cos\left(\frac{\pi}{6}\right)X+\sin\left(\frac{\pi}{6}\right)Y,\nonumber \\
A_{2}=B_{2}=C_{2} & = & \cos\left(\frac{\pi}{12}\right)X-\sin\left(\frac{\pi}{12}\right)Y,\label{eq:MerminMeasurements}
\end{eqnarray}
where $X$ and $Y$ are Pauli operators. We have $H\left(A_{1}\otimes B_{2} \otimes C_{2}\right)=H\left(A_{2}\otimes B_{1} \otimes C_{2}\right)=H\left(A_{2}\otimes B_{2} \otimes C_{1}\right)=0$, but at the same time $H\left(A_{1}\otimes B_{1} \otimes C_{1}\right)=1$. This achieves maximal algebraic violation of (\ref{quantum}).

Note that the derivation of (\ref{eq:EntropicMermin}) holds not only for the entropic function $\delta$, but for any distance with the associativity property. For instance, applying the generalisation of the co-variance distance \cite{distance1,distance2}, $\delta\left(A_{1},A_{2},A_{3}\right)=1-\left\langle A_{1}\cdot A_{2}\cdot A_{3}\right\rangle $ to (\ref{eq:EntropicMermin}) gives the original tripartite Mermin inequality \cite{Mermin}
\begin{eqnarray}
& &\left\langle A_{1}\cdot B_{2}\cdot C_{2}\right\rangle+\left\langle A_{2}\cdot B_{1}\cdot C_{2}\right\rangle+\left\langle A_{2}\cdot B_{2}\cdot C_{1}\right\rangle \nonumber \\ &-&\left\langle A_{1}\cdot B_{1}\cdot C_{1}\right\rangle \leq 2.
\end{eqnarray}

%%%%%%%%%%%%%%%%%%%%%%%%%%%%%%%%%%%%%%%%%%%%%%%%%%

\emph{Multipartite information-theoretic Bell inequality.} One way to extend the previous result to more than three parties is to follow the approach proposed in \cite{multiGHZ}, where it was shown that the genuine multipartite nonlocality without inequalities, the so called Greenberger-Horne-Zeilinger (GHZ) paradox \cite{GHZ}, can be obtained if one allows for more than two measurements per observer. We will focus on the even number of qubits, which requires three measurements per observer. However this reasoning should be applicable to the odd number of qubits as well, perhaps with the need for more than three local measurements \cite{multiGHZ}.

Let us consider an even number $N$ of observers sharing a multipartite system. Each observer measures three randomly chosen binary $\pm 1$ observables $M_{i}^{(j)}$, where $i=1,2,3$ labels measurements and $j=1,\dots,N$ labels observers. We use the associativity, symmetry and the triangle inequality of (\ref{d}) to obtain 
\begin{eqnarray}
& &\delta(M_{1}^{(1)},M_{1}^{(2)},M_{1}^{(3)},M_{1}^{(4)},\ldots,M_{1}^{(N)}) \leq \nonumber \\
& &\delta(M_{1}^{(1)},M_{2}^{(2)},M_{3}^{(3)},M_{3}^{(4)},\ldots,M_{3}^{(N)}) +  \nonumber \\
& &\delta(M_{3}^{(1)},M_{1}^{(2)},M_{2}^{(3)},M_{3}^{(4)},\ldots,M_{3}^{(N)}) +  \nonumber \\
& &\delta(M_{3}^{(1)},M_{3}^{(2)},M_{1}^{(3)},M_{2}^{(4)},\ldots,M_{3}^{(N)}) + \ldots + \nonumber \\
& &\delta(M_{2}^{(1)},M_{3}^{(2)},M_{3}^{(3)},M_{3}^{(4)},\ldots,M_{1}^{(N)}) + \nonumber \\
& &\delta(M_{2}^{(1)},M_{2}^{(2)},M_{2}^{(3)},M_{2}^{(4)},\ldots,M_{2}^{(N)}). \label{multi}
\end{eqnarray}
The term on the left hand side contains only measurements $M_{1}^{(j)}$, whereas the first $N$ terms on the right are cyclic permutations of one measurement $M_{1}^{(j)}$, one measurement $M_{2}^{(j)}$ and $N-2$ measurements $M_{3}^{(j)}$. The remaining term on the right contains only measurements $M_{2}^{(j)}$.

The derivation is as follows. We start with the multipartite distance $\delta(M_{1}^{(1)},M_{1}^{(2)},M_{1}^{(3)},M_{1}^{(4)},\ldots,M_{1}^{(N)})$ and apply the triangle inequality (together with symmetry and associativity) to obtain
\begin{eqnarray}
& &\delta(M_{1}^{(1)},M_{1}^{(2)},M_{1}^{(3)},M_{1}^{(4)},\ldots,M_{1}^{(N)}) \leq \nonumber \\
& &\delta(M_{1}^{(1)},M_{2}^{(2)},M_{3}^{(3)},M_{3}^{(4)},\ldots,M_{3}^{(N)}) + \nonumber \\
& &\delta(M_{1}^{(2)},\ldots,M_{1}^{(N)},M_{2}^{(2)},M_{3}^{(3)},\ldots,M_{3}^{(N)}).
\end{eqnarray}
The term on the left and the first term on the right hand side correspond to measurable quantities in the inequality (\ref{multi}), whereas the second term on the right cannot be experimentally verified. Therefore, we repeat the same procedure as before and expand the last term as
\begin{eqnarray}
& &\delta(M_{1}^{(2)},\ldots,M_{1}^{(N)},M_{2}^{(2)},M_{3}^{(3)},\ldots,M_{3}^{(N)}) \leq \nonumber \\
& &\delta(M_{3}^{(1)},M_{1}^{(2)},M_{2}^{(3)},M_{3}^{(4)},\ldots,M_{3}^{(N)}) + \nonumber \\
& &\delta(M_{1}^{(3)},\ldots,M_{1}^{(N)},M_{2}^{(2)},M_{2}^{(3)},M_{3}^{(1)},M_{3}^{(3)}).
\end{eqnarray}
Again, we generated a term that is observable and an additional term that requires further application of the triangle inequality. One can easily notice the following pattern. After $k$ repetitions of the above procedure one generates the measurable term 
\begin{equation}
\delta(M_{3}^{(1)},\ldots,M_{3}^{(k-1)},M_{1}^{(k)},M_{2}^{(k+1)},M_{3}^{(k+2)},\ldots,M_{3}^{(N)})
\end{equation}
and the non-measurable term of the form
\begin{eqnarray}
& &\delta(M_{1}^{(k+1)},\ldots,M_{1}^{(N)},M_{2}^{(2)},\ldots,M_{2}^{(k+1)},M_{3}^{(2)},\ldots,\nonumber \\ 
& &M_{3}^{(k)}, M_{3}^{(k+2)},\ldots,M_{3}^{(N)})
\end{eqnarray}
if $k$ is odd, or
\begin{equation}
\delta(M_{1}^{(k+1)},\ldots,M_{1}^{(N)},M_{2}^{(2)},\ldots,M_{2}^{(k+1)},M_{3}^{(1)},M_{3}^{(k+1)})
\end{equation}
if $k$ is even. Finally, after $k=N-1$ repetitions (we remember that $N$ is even) we obtain almost all the right hand side terms of (\ref{multi}), except the last two, and an additional non-measurable term. This term can be expanded into two missing measurable ones
\begin{eqnarray}
& &\delta(M_{1}^{(N)},M_{2}^{(2)},\ldots,M_{2}^{(N)},M_{3}^{(2)},\ldots,M_{3}^{(N-1)}) \leq \nonumber \\
& &\delta(M_{2}^{(1)},M_{3}^{(2)},M_{3}^{(3)},\ldots,M_{3}^{(N-1)},M_{1}^{(N)}) + \nonumber \\
& &\delta(M_{2}^{(1)},M_{2}^{(2)},M_{2}^{(3)},M_{2}^{(4)},\ldots,M_{2}^{(N)}),
\end{eqnarray}
which ends the derivation.

The above inequality can be maximally violated in quantum mechanics by the $N$-partite GHZ state $|GHZ\rangle_N=(|0\ldots 0\rangle + |1\ldots 1\rangle)/\sqrt{2}$ and for $\delta(M_{1}^{(1)},M_{1}^{(2)},M_{1}^{(3)},M_{1}^{(4)},\ldots,M_{1}^{(N)})= H(M_{1}^{(1)}\otimes M_{1}^{(2)}\otimes \ldots \otimes M_{1}^{(N)})$, etc. In this case the local measurements are $M_{i}^{(j)} = \cos \alpha_i X + \sin \alpha_i Y$, where $\alpha_1 = \frac{\pi}{2N}$, $\alpha_2 = 0$ and $\alpha_3= -\frac{\pi}{2N(N-2)}$. The choice of the angles stems from the following observation
\begin{eqnarray}
& & \left(M_{i}^{(1)}\otimes \ldots \otimes M_{j}^{(N)}\right) \frac{|0\ldots 0\rangle + |1\ldots 1\rangle}{2} = \\
& &  \frac{e^{-i(\alpha_i+\ldots +\alpha_j)}|0\ldots 0\rangle + e^{i(\alpha_i+\ldots +\alpha_j)}|1\ldots 1\rangle}{2} = |\overline{GHZ}\rangle_N,  \nonumber
\end{eqnarray} 
where the overlap $\langle GHZ |\overline{GHZ}\rangle_N = \cos (\alpha_i + \ldots + \alpha_j)$. In every case, except $H(M_{1}^{(1)}\otimes M_{1}^{(2)}\otimes \ldots \otimes M_{1}^{(N)})$, the overlap is one and the corresponding entropy is zero. On the other hand, the entropy on the left hand side of (\ref{multi}) is one because the overlap is zero.

%%%%%%%%%%%%%%%%%%%%%%%%%%%%%%%%%%%%%%%%%%%%%%%%%%

\emph{Information-theoretic state-independent contextuality.} Contextuality is a form of nonclassicality that is more general than nonlocality. In this case a spatial separation of measurements is not necessary and all of them can be performed on a single localised system. The crucial assumption is based on a classical intuition that the outcome of one measurement does not depend on what other compatible (non-disturbing) measurement is performed at the same time. This assumption is known as non-contextuality and systems violating it are called contextual. Interestingly, in quantum theory contextuality can be exhibited by any state of the system with the dimension larger than two, whereas nonclassicality in nonlocal scenarios can be exhibited only by entangled states. 

We will now consider an entropic version of the state-independent contextuality proof commonly known as Peres-Mermin square \cite{PM1,PM2,PM3}. Consider nine $\pm1$ observables that can be measured on a single system. Due to compatibility relations these measurements can be performed in the following triples: $\{A,a,\alpha\},\{B,b,\beta\},\{C,c,\gamma\},\{A,B,C\},\{a,b,c\},\{\alpha,\beta,\gamma\}$. The classical reasoning, based on the non-contextuality assumption and on the assumption that outcomes exist prior measurements, implies that for measurable products $q_1=A\cdot a\cdot \alpha$, $q_2=B\cdot b\cdot \beta$, $q_3=C\cdot c\cdot \gamma$, $q_4=A\cdot B\cdot C$, $q_5=a\cdot b\cdot c$, and $q_6=\alpha\cdot \beta\cdot\gamma$ one has $\prod_{i=1}^6 q_i =1$. 

\begin{figure}[t]
    \begin{center}
    	\includegraphics[width=1.0\columnwidth,trim=4 4 4 4,clip]{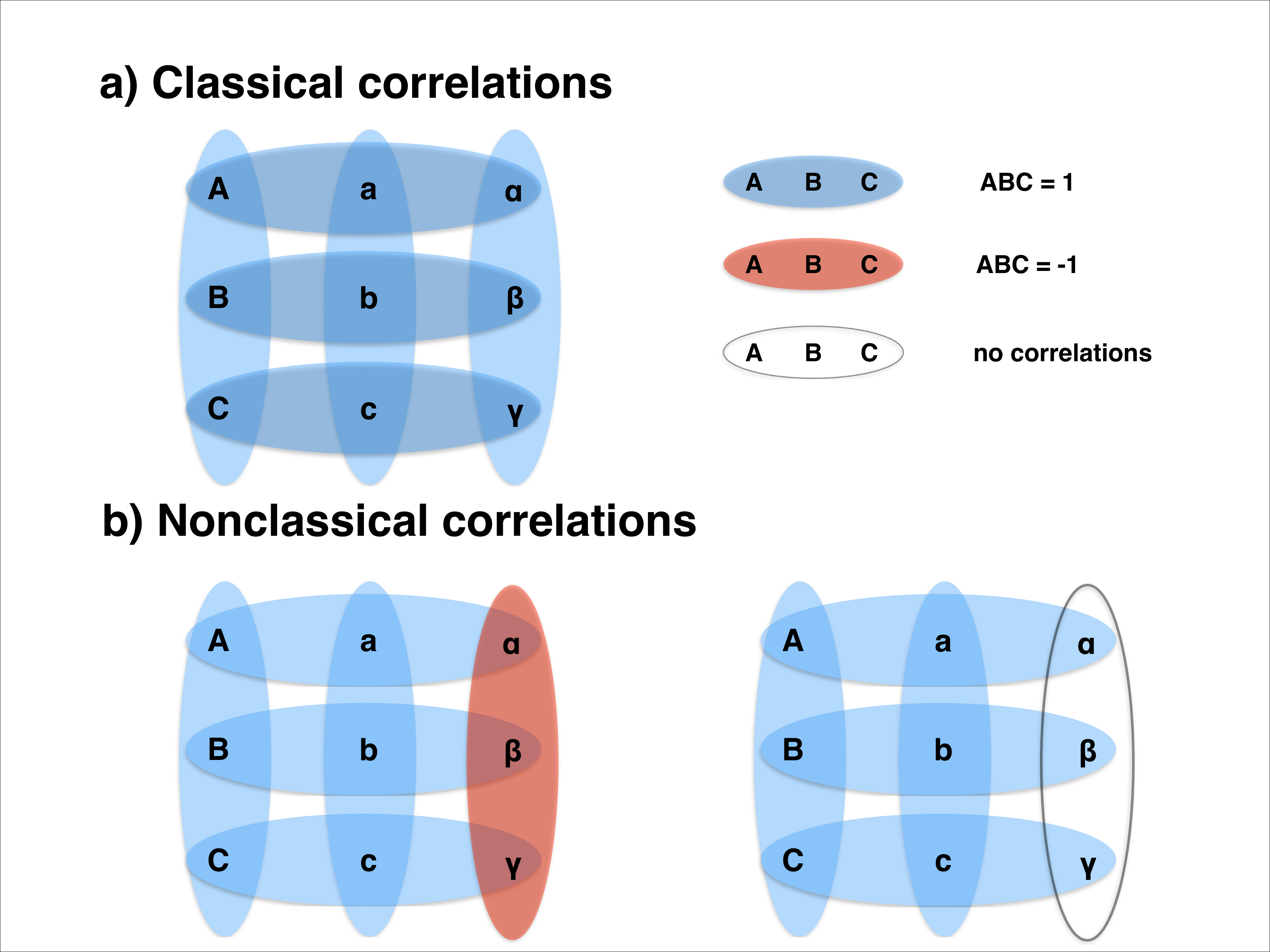}
        \caption{Tripartite correlations between local observables constituting the Peres-Mermin square \cite{PM1,PM2,PM3}. Classical correlations (\emph{a}) imply that $(Aa\alpha)(Bb\beta)(Cc\gamma)(ABC)(abc)(\alpha\beta\gamma)=1$. However, for nonclassical contextual correlations (\emph{b} left) the product can give $-1$, for example $\alpha\beta\gamma=-1$. In quantum mechanics one can find such non contextual observables for which these nonclassical correlations do not depend on a state of a system. They are entropically indistinguishable from classical ones, however, after mixing them with the classical ones one obtains a distribution that can be detected by entropic inequalities (\emph{b} right) since $H(\alpha\cdot \beta \cdot \gamma)=1$}
        \label{fig3}
    \end{center}
\end{figure}

However, in nonclassical theories one can observe that $\prod_{i=1}^6 q_i =-1$ (see Fig. \ref{fig3}). In quantum theory this is achieved by a set of two-qubit measurements  $A=X\otimes\openone$, $a=\openone\otimes X$, $\alpha=X\otimes X$, $B=\openone\otimes Y$, $a=Y\otimes \openone$, $\beta=Y\otimes Y$, $C=X\otimes Y$, $c=Y\otimes X$, $\gamma=Z\otimes Z$, where $X$, $Y$ and $Z$ are Pauli operators \cite{PM1,PM2,PM3}.

Next, we use the associativity, symmetry and the triangle inequality to derive: 
\begin{eqnarray}
& &\delta(\alpha,\beta,\gamma) \leq \delta(A,a,\alpha) + \delta(A,a,\beta,\gamma) \leq \label{dsi} \\
& & \delta(A,a,\alpha) + \delta(B,b,\beta) + \delta(A,a,B,b,\gamma) \leq \nonumber \\
& & \delta(A,a,\alpha) + \delta(B,b,\beta) + \delta(A,B,C) + \delta(C,a,b,\gamma) \leq \nonumber \\
& & \delta(A,a,\alpha) + \delta(B,b,\beta) + \delta(A,B,C) + \delta(a,b,c) +\delta(C,c,\gamma). \nonumber
\end{eqnarray}
The corresponding entropic inequality is of the form
\begin{eqnarray}
& &H(\alpha\cdot\beta\cdot\gamma) \leq H(A\cdot a \cdot \alpha) + H(B\cdot b\cdot \beta) \nonumber \\ &+& H(A\cdot B\cdot C) + H(a\cdot b\cdot c) + H(C\cdot c \cdot \gamma). \label{esi}
\end{eqnarray}

For the above quantum observables on a localised two-qubit system one finds that $q_1=q_2=\dots=q_5=1$ and $q_6=-1$ for any quantum state. This distribution of outcomes does not violate the inequality (\ref{esi}), however if we follow the method of Ref. \cite{Chaves} and equally mix it with the classical distribution $q_1=\dots=q_6=1$ we get a distribution allowing for the maximal violation (see Fig. \ref{fig3} b right).

As in the case of the multipartite nonlocality, the derivation of (\ref{dsi}) holds not only for the function $\delta$, but for any distance with the associativity property. Applying $\delta\left(A_{1},A_{2},A_{3}\right)=1-\left\langle A_{1}\cdot A_{2}\cdot A_{3}\right\rangle $ to the inequality (\ref{dsi}) gives the non-contextuality inequality by Cabello \cite{Cabello}, which is violated by any quantum state
\begin{eqnarray}
& & \langle A\cdot a \cdot \alpha\rangle + \langle B\cdot b\cdot \beta\rangle + \langle A\cdot B\cdot C\rangle \nonumber \\ &+& \langle a\cdot b\cdot c\rangle + \langle C\cdot c \cdot \gamma\rangle - \langle\alpha\cdot\beta\cdot\gamma \rangle \leq 4.
\end{eqnarray}

%%%%%%%%%%%%%%%%%%%%%%%%%%%%%%%%%%%%%%%%%%%%%%%%%%

\emph{Discussion.} We have derived entropic inequalities to test the multipartite nonlocality and the state-independent contextuality using the multipartite information-theoretic distance and a finite number of measurements. In both cases, quantum mechanics allows for maximal violation of these inequalities, which corresponds to the maximal nonclassical behaviour detectable by such a test. On the other hand, the bipartite quantum nonlocality was unable to maximally violate entropic inequalities based on a finite number of measurements \cite{BC}. This brings us to an interesting analogy. It was shown (see for example \cite{Mermin}) that correlation-based Bell inequalities admit maximal quantum violation only in the case of multipartite systems. Here we show that the same is true for multipartite entropic Bell inequalities.

Another important observation is the fact that for the multipartite nonlocality we were able to find a state and measurements leading to a direct violation, whereas in the case of the state-independent contextuality we had to mix the measured nonclassical data with a classical probability distribution in order to observe the violation. We attribute this to the state-independence property. The state-independent contextuality is a property of measurements, not the property of states. In order to obtain a direct violation of an entropic inequality one would have to look for a product of some observables for which the entropy in any state is larger than entropies of other products. Although we do not provide a proof, we speculate that such observables do not exist.

We also re-derived two known inequalities. This suggests that the multipartite information-theoretic distance approach is a more general treatment of multipartite nonclassical correlations than the standard approach with correlation functions. Note, that this idea has been already proposed for bipartite nonclassical correlations \cite{distance1}. However, to fully prove it one needs to show that all multipartite Bell inequalities and non-contextuality inequalities can be derived from some multipartite distance.

There are several open problems that require further investigation: (i) extension to nonlocality of an odd number of qubits (ii) finding an information-theoretic distance suitable to investigate multipartite nonlocality of higher dimensional quantum systems (iii) derive multipartite monogamy relations from the properties of information-theoretic distances, perhaps using ideas in \cite{distance1}.

%%%%%%%%%%%%%%%%%%%%%%%%%%%%%%%%%%%%%%%%%%%%%%%%%%

\emph{Acknowledgements.} S.R. was supported by Canada’s NSERC, 
MPrime, CIFAR, and CFI and IQC. 
P. K. and D. K. are supported by the Foundational Questions Institute (FQXi) and by the National Research Foundation and Ministry of Education in Singapore.

%%%%%%%%%%%%%%%%%%%%%%%%%%%%%%%%%%%%%%%%%%%%%%%%%%

%%%%%%%%%%%%%%%%%%%%%%%%%%%%%%%%%%%%%%%%%%%%%%%%%%

\end{document}